# The effect of different annealing cooling rate on electrical and mechanical properties of TiO$_2$ thin films

Shokoufeh. Asalzadeh[1, *], Kiomars. Yasserian[1]

[1]Department of Physics, Islamic Azad University, Karaj Branch, Iran

[*]sasalzad@hawk.iit.edu

Abstract

This paper presents the effect of different cooling rate after annealing on structural and electrical properties of $TiO_2$ thin films. Titanium dioxide thin film was deposited on a silicon substrate using DC magnetron technique. After annealing $TiO_2$ thin films at 600°C, to investigate the effect of different cooling rates on $TiO_2$ thin films, samples were cooled down from 600°C to the room temperature under 3 different rates: 2°C/min, 6°C/min and 8°C/min. The Surface morphology, crystal structure, and electrical property of samples were characterized by atomic force measurement (AFM), X-ray diffraction (XRD) and Four-point probe (FPP) technique. It is found that the rate of decreasing temperature after annealing can affect the morphology structure and electrical resistivity of titanium dioxide. Sample with 2°C/m has the biggest grain size, and more electrical resistivity while the smallest grain size and lowest electrical resistivity belong to sample with 8°C/m cooling rate.

1.INTRODUCTION:

$TiO_2$ as one of the most thermally and chemically stable semiconductor has been attractive for many researchers [1-6]. Due to its interesting chemical, electrical and optical properties [7,8] it has been widely used in many application like solar cells [9], photocatalysis [10], gas sensor [11]. Also, because of its low electrical resistivity, $TiO_2$ is an interesting material for microelectronic devices [12].The $TiO_2$ thin film can be fabricated using various methods, like chemical vapor deposition (CVD), ion beam deposition, sol gel dip, plasma enhance chemical vapor deposition, RF magnetron sputtering, and DC magnetron sputtering [13,15]

Since DC magnetron sputtering is a more controllable deposition technique that provides more uniform coated thin layers [16], in our study we fabricated $TiO_2$ thin film samples using this method. $TiO_2$ exists in 3 different crystal phases: rutile (tetragonal), anatase (tetragonal), and brookite (orthorhombic) [17,18]. Rutile compared to other phases has more stability while anatase and brookite are metastable and can be easily transformed to anatase by heating [19,20]. Annealing at 600 °C creates rutile phase in annealed $TiO_2$ [21].

Some recent studies reported the effect of heat treatment and annealing time and temperature on structural, mechanical, and optical properties of $TiO_2$ thin films [22-25]. Also, other research has studied light harvesting and efficiency in solar cells [26] and electrical properties of $TiO_2$ [27]. According to [28] different annealing rate (ramped from room temperature to 450°C) can affect the surface roughness of $TiO_2$ thin films. Authors of [28] showed that the slowest rate has the lowest roughness and higher size of islands.

Another study investigated the effect of ultra-fast annealing on electrical properties of $TiO_2$ Polycrystal thin films. They annealed the $TiO_2$ at 550°C (with 10K/min ramp) for 1 hour and at 460 °C (with 300 K/min ramp) under oxygen treatment and fast cooling after annealing. The total ultra-fast treatment including heating and cooling was done in 5 minutes. Ultra-fast annealed $TiO_2$ showed lower resistivity Compared to the resistivity of the standard annealed $TiO_2$[29].

In an effort to control the semi conductivity of $TiO_2$, we modified the annealing process by controlling the cooling rate after the annealing. We showed different cooling rate after annealing $TiO_2$ thin films changed the electrical properties. In Chapter 2, the sample preparation procedure is presented followed by the experimental results and discussions presented in Chapter 3 and the conclusions in Chapter 4. The intensity of rutile peak for the samples with different cooling rates after annealing was varied and it was observed that it affected the electrical resistivity and conductivity. TiO₂ thin films with different ramping rate have different grain sizes and roughness which leads to changes in electrical properties.

## 2. EXPERIMENTAL:

TiO₂ thin films were coated on silicon substrate by DC magnetron sputtering technique at room temperature.

Before deposition, the 1cm ×1cm silicon substrates were cleaned in an ultrasonic bath for 10 minutes. The argon (99.99% pure) as the sputtering gas and high purity oxygen (99.99%) were used with the composition from pure argon to pure oxygen. Ti target and substrate distance was fixed at 35mm. Sputtering chamber evacuated to $p = 2 \times 10^{-5}$ and the pressure of working gas kept at $p = 2 \times 10^{-2}$ Torr, with discharge current and electric discharge potential of 200 mA and 500 V, respectively. After deposition, by using Dektak3 surface profile measurement system, 35 nm thickness was measured for all samples.

The TiO₂ thin films were annealed at 600°C temperatures [30, 31, 32] for 10 minutes, the temperature was decreased from 600°C to the room temperature under three different rates of 2°C/min, 6°C/min and 8°C/min. XRD, AFM, and FPP measurement methods were employed to analyze the effect of different cooling rates on electrical and morphological properties of TiO₂ thin films, presented in the ensuing Chapter.

## 3. EXPERIMENTAL RESULTS AND DISCUSSIONS: 3.1. XRD RESULTS:

Figure 1 shows the XRD results of TiO₂ thin films cooled down at different rate

(2°C/min, 6°C/min, and 8°C/min) after annealing. X-Ray pattern measured by CUKA 1 source and 1.5 Å wavelength, scanning 2θ in range of 30° to 75°. The main peak around 69° is attributed to the Silicon (400) substrate (reference code:00-033-1381).

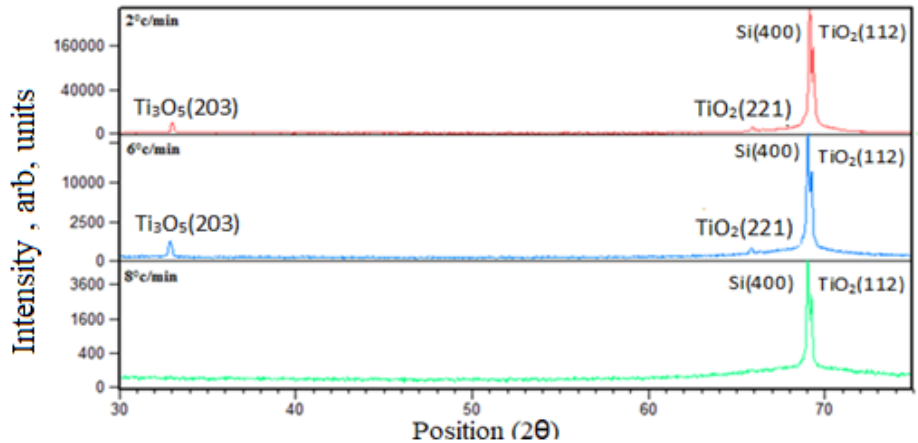

Fig.1.XRD spectra of $TiO_2$ thin film cooling down to the room temperature after annealing at 600°C with a) 2°C/min b) 6°C/min c) 8°C/min rate

As it was expected annealing $TiO_2$ at 600°C produce the rutile phase [21]. The XRD spectra indicates that all the samples consist of $TiO_2$ rutile phase peak with (112) preferred orientation as the shoulder at the right side of Si peak.

For the $TiO_2$ thin film cooling with 8°C/m, this phase was observed at 2θ=69.22° with lower intensity and tetragonal structure. With decreasing cooling rate, the intensity of the rutile phase increased. Obvious increased intensity of the $TiO_2$ peak rutile phase (112) (reference code: 01-084-1284) peak for 6° C/m sample is observed compared to 8°C/m sample. Also, XRD data shows multi crystal structure including $Ti_3O_5$(203) (reference code 01-082-1137) with monoclinic structure, and $TiO_2$(221) (reference code: 01-084-1284) rutile phase with tetragonal structure started to appear for this sample.

For 2°C/m sample, rutile phase peak (112) has a higher intensity compared to two other samples. The intensity enhancement can be seen in Ti$_3$O$_5$(203) and TiO$_2$(221) peaks as well. This result shows that slowing down the cooling rate after annealing TiO$_2$ creates better crystallization and phase stability.

The average grain size can be calculated by XRD data and the Scherrer Equation [33]:

$$D = \frac{k\lambda}{\beta \cos(\theta)} \quad (1)$$

where D is the grain size in nm, k is the shape constant (0.9), β is the peak full width of half maximum (FWHM), λ is the X-ray wavelength (1.54 Å), and θ is the diffraction angle.

The results indicated that by speeding up the cooling rate, grain sizes were decreased. Table 1 represents the measured grain sizes of samples.

| Sample | Position (2θ) | FWHM(θ) | Crystal size (nm) |
|---|---|---|---|
| 8°C/m | 69.04(112) | 1.14 | 0.93 |
| 6°C/m | 69.22(112) | 3.43 | 0.82 |
|  | 65.77(221) | 13.75 | 0.83 |
|  | 32.83(203) | 9.16 | 1.19 |
| 2°C/m | 69.32(112) | 1.14 | 0.93 |
|  | 65.88(221) | 9.16 | 1.09 |
|  | 32.96(203) | 6.87 | 1.24 |

Table 1. Crystal size of TiO$_2$ with different ramp rates (2°C/m, 6°C/m and 8°C/m)

## 3.2. AFM RESULTS:

Surface morphology and roughness of TiO$_2$ thin films with different cooling rates (2°C/m, 6°C/m, and 8°C/m) after annealing, were evaluated by AFM imaging. Figure 2 presents the AFM images of the TiO$_2$ thin films with a ramping rate of a) 2°C/m b) 6°C/m and c) 8°C/m. Increasing the grain size with lowering the cooling rate obtained from XRD results, is in consistency with AFM results.

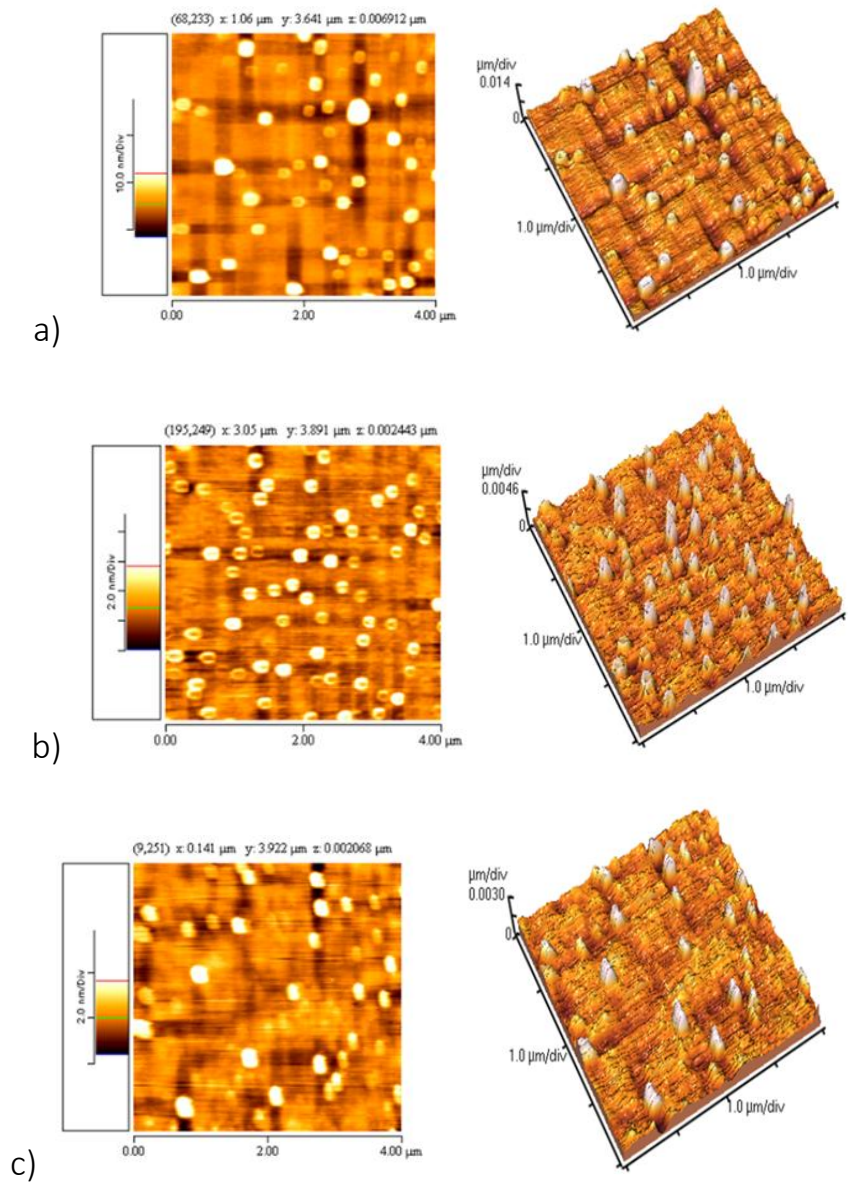

Fig .2. 2D and 3D AFM images of TiO$_2$ thin film cooling down to the room temperature after annealing at 600°C with a) 2°C/m b) 6°C/m c) 8°C/m rate

In AFM analysis, surface roughness is presented by root means square (Rq) [34-35].

According to the AFM results, surface roughness was measured as 1.862nm, 2.002nm to 2.294 nm for 8°C/m, 6°C/m to 2°C/m cooling rates, respectively. It is observed that when the cooling rate decreases, grain size and roughness increased. Densification in lower rates leads to higher grain size. This effect can be seen in Figure 3.

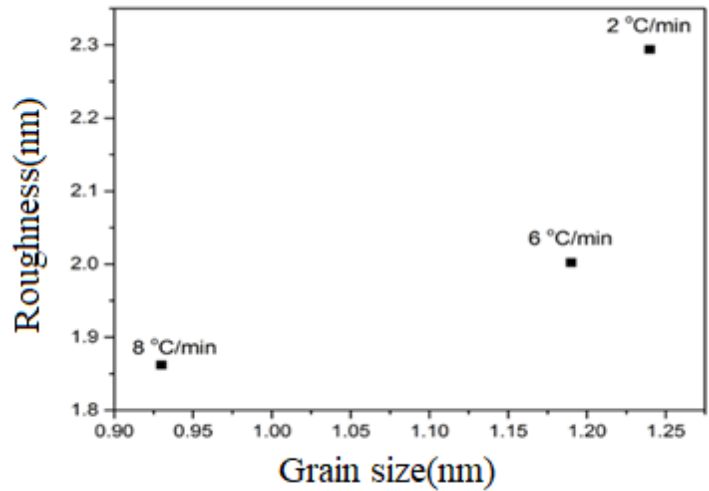

Fig.3. Grain size Vs. the roughness of TiO$_2$ thin film cooling with different ramp rate (2,6 and 8°C/m)

3.3. FPP RESULTS:

The FPP technique was used to measure the resistivity of different TiO$_2$ thin films.

Surface resistivity ($\rho_s$) can be calculated by Equation (2) [36]:

$$\rho_s = \left(\frac{\pi}{\ln 2}\right) f \left(\frac{R_1+R_2}{2}\right) \quad (2)$$

F is Vander Paw factor that depends on the position of electrical connection can be calculated by Equation (3):

$$f = 1 - \frac{\ln 2}{2}\left[\frac{R_1 - R_2}{R_1 + R_2}\right]^2 \quad (3)$$

electrical resistivity can be calculated by $\rho = t\, \rho_s$, t is the thickness of the thin film, which is 35nm in this study.

Equation (4) describes Electrical conductivity ($\sigma$) that has a reciprocal relationship with the resistivity ($\rho$).

$$\sigma = \frac{1}{\rho} \quad (4)$$

As Equation (4) shows that when resistivity increases, the conductivity decreases. Figure 4 illustrates measured resistivity and calculated conductivity of $TiO_2$ thin films with different cooling rates. Increasing the cooling rate leads to decreased surface roughness and increased conductivity. Resistivity measured 80 µΩ, 125 µΩ, 185 µΩ for the TiO2 thin films with the ramping rate of 8° C/m, 6° C/m and 2° C/m respectively.

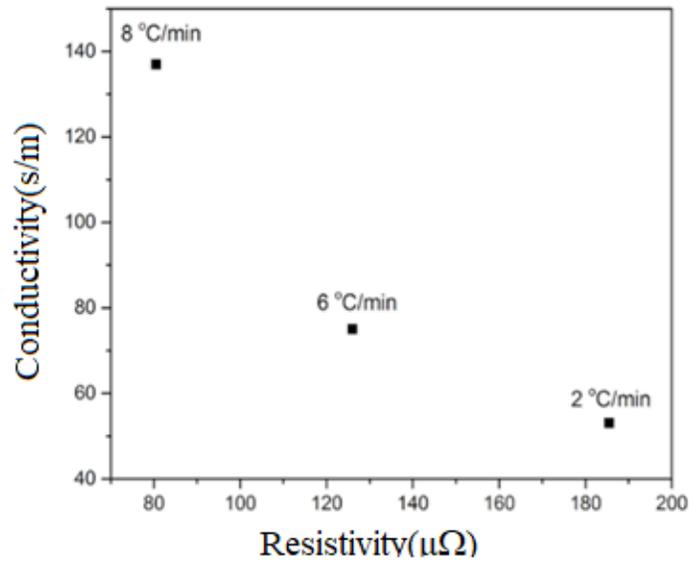

Fig .4. Resistivity Vs. conductivity of $TiO_2$ thin film with 2°C/m,6°C/m,8°C/m of cooling rate

According to the AFM and XRD data, the Lowest ramping rate has the highest roughness and grain size, improved crystallite, higher oxidation phase, and highest resistivity of 185µΩ. Compared to other samples, the samples cooled with 2°C/m rate demonstrated the most improved crystallite and semiconducting property which lead to the highest resistivity. In addition, increasing grain size leads to decreasing grain boundary and hence conductivity [37]. The clump shape of grains and valleys between the clumped grains reduce the movement of charge careers.

lower oxidation phase, weaker semiconducting properties and smaller grain sizes of 6°C/m and 8°C/m sample lead to the decrease of the electrical resistivity. The decrease in grain size enhances the electron mobility and consequently increases the conductivity. 8°C/m sample has the lowest grain size and lowest resistivity of 80 µΩ , 6°C/m rate has 123 µΩ resistivities.

## 4.CONCLUSION:

In this work the influence of different cooling rates (2° C/m, 6°C/m and 8°C/m) after the annealing process on morphological and electrical properties of $TiO_2$ thin film coated on silicon substrate fabricated by DC magnetron technique was investigated. It has been observed that the slowing the ramping rate increased the grain size, roughness, and conductivity of $TiO_2$ thin films. 2°C/m rate has the largest grain size, roughness, semiconducting properties, and improved crystallite and resistivity of 185 µΩ, while the highest cooling rate (8°C/m) has the lowest grain size and resistivity of 80 µΩ and consequently the highest conductivity. A resistivity of 123 µΩ was measured for the for 6°C/m sample.

.